\documentclass[aps,prl,twocolumn,groupedaddress]{revtex4}
\usepackage[dvips]{graphicx,color}
\usepackage{amsmath,amssymb, booktabs}

\begin{document}

\newcommand{\etal}      {{\it et~al.}}

\title{Search for Two-Photon Interaction with Axionlike Particles\\Using High-Repetition Pulsed Magnets and Synchrotron X Rays}

\author{T. Inada,$^1$ T. Yamazaki,$^1$ T. Namba,$^1$ S. Asai,$^2$ T. Kobayashi,$^1$ K. Tamasaku,$^3$ Y. Tanaka,$^4$ Y. Inubushi,$^3$ K. Sawada,$^3$ M. Yabashi,$^3$ T. Ishikawa,$^3$ A. Matsuo,$^5$ K. Kawaguchi,$^5$ K. Kindo,$^5$ H. Nojiri$^6$}
\affiliation{$^1$International Center for Elementary Particle Physics, The University of Tokyo, 7-3-1 Hongo, Bunkyo-ku, Tokyo 113-0033, Japan}
\affiliation{$^2$Department of Physics, Graduate School of Science, The University of Tokyo, 7-3-1 Hongo, Bunkyo-ku, Tokyo 113-0033, Japan}
\affiliation{$^3$RIKEN SPring-8 Center, 1-1-1 Kouto, Sayo-cho, Sayo-gun, Hyogo 679-5148, Japan}
\affiliation{$^4$Graduate School of Material Science, University of Hyogo, Kamigori, Hyogo 678-1297, Japan}
\affiliation{$^5$The Institute for Solid State Physics, The University of Tokyo, 5-1-5 Kashiwanoha, Kashiwa-shi, Chiba 277-8581, Japan}
\affiliation{$^6$Institute for Materials Research, Tohoku University, 2-1-1 Katahira, Aoba-ku, Sendai 980-8577, Japan}

\date{\today}

\begin{abstract}
We report on new results of a search for two-photon interaction with axionlike particles (ALPs).
The experiment was carried out at a synchrotron radiation facility using a ``light shining through a wall (LSW)'' technique.
For this purpose, we have developed a novel pulsed-magnet system, composed of multiple racetrack-magnets and a transportable power supply.
It produces fields of about 10 T over 0.8 m with a high repetition rate of 0.2 Hz and yields a new method of probing vacuum with high intensity fields.
The data obtained with a total of 27,676 pulses provide a limit on the ALP-two-photon coupling constant that is more stringent by a factor of 5.2 compared to a previous x-ray LSW limit for the ALP mass $\lesssim0.1$ eV.
\end{abstract}

\maketitle

Photons are one of the most fundamental objects in physics, and their direct interaction does not take place in vacuum because of their electrical neutrality.
However, a nonlinear effect of quantum electrodynamics (QED) predicts the two-photon interaction (TPI), intermediated by a virtual electron-positron loop~\cite{qed1}.
This effect is currently being studied in many experiments by using a scattering method~\cite{gamgam1, gamgam2, gamgam3, gamgam4} or by measuring vacuum magnetic birefringence (VMB)~\cite{cm, pvlas1, pvlas2, bmv}.
While this QED cross section is quite small~\cite{qed2, qed3}, the existence of possible new scalar or pseudoscalar particles could provide additional contributions via the Primakoff effect~\cite{primakov1, primakov2}.
The first observation of TPI with an elementary scalar field was provided by the recent discovery of Higgs boson, decaying with $H \rightarrow \gamma \gamma$~\cite{atlas, cms}.
Another well-known example for a pseudoscalar field is a neutral pion decay $\pi^{0} \rightarrow \gamma \gamma$~\cite{pdg}.
Therefore, searches for new (pseudo-)scalar bosons with a small mass like axions~\cite{axion1, axion2, cp1, cp2, cp3} and axionlike particles (ALPs)~\cite{susy, string1, string2, csm} are important in the context of particle physics~\cite{ringwald},
and their contribution to TPI may also reveal a new aspect of photons.

\vspace{-1.2mm}
TPI with an axionlike pseudoscalar field $\phi_{a}$ is given by the Lagrangian $L_{a\gamma\gamma}=g_{a\gamma\gamma}\vec{E}\cdot \vec{B}\,\phi_{a}$,
with $g_{a\gamma\gamma}$ their coupling constant and $\vec{E}\cdot\vec{B}$ the odd-parity product of an electric and magnetic fields.
Up to now, a large number of laboratorial searches for ALPs have been carried out with a ``light shining through a wall (LSW)'' technique~\cite{lsw_summary},
where $\vec{E}$ is provided as real photons from a light source and $\vec{B}$ as virtual photons from an external magnetic field, so that the field direction coincides with the polarization of the incident light.
The generated ALPs pass through a beam dump that blocks the unconverted photons.
Some of the ALPs then reconvert into detectable photons by an inverse process in a second magnet.

Most of the previous experiments used optical lasers as the light source~\cite{laser1, laser2, pulse1, laser3, laser4, pulse2, laser5, laser6, laser7, alps, osqar}.
However, unlike invisible axion models~\cite{model1, model2, model3, model4}, ALP mass $m_{a}$ and $g_{a \gamma \gamma}$ are not bound with each other.
Thus, new parameter spaces have been searched with various photon energies (see \cite{lsw_summary2} and references therein).
To probe a higher mass region than that of laser experiments,
the use of synchrotron x-rays was proposed~\cite{x_proposal1, x_proposal2}.
The photon energy is typically higher by 3--4 orders of magnitude compared to optical photons.
The first x-ray LSW search was carried out at the European Synchrotron Radiation Facility (ESRF) with superconducting magnets, significantly extending the limit on $g_{a\gamma\gamma}$ up to around 1 eV~\cite{esrf}.
While searches for ALP flux from the Sun have also probed this mass region~\cite{cast1, cast2, cast3, cast4, sumico1, sumico2, sumico3},
the flux estimation inevitably relys on a solar model~\cite{evade1, evade2, evade3, evade4, evade5, evade6} and its complex magnetic activity~\cite{solar_mag}.
ALP interaction with the solar magnetic field has been implied by its x-ray spectra, coronal heating, and cyclic luminosity variation for ALP masses around 20 meV~\cite{zioutas1, zioutas2, rusov1, rusov2},
showing the importance of complementary searches using terrestrial and extra-terrestrial x-ray sources~\cite{esrf}.

In this Letter, we describe a new method of searching for ALPs at a synchrotron radiation (SR) facility by applying highly repetitive pulsed fields.
The key technology of the experiment is a magnet system.
Its requirements are (i) field direction transverse to the light propagation, (ii) high field intensity, (iii) large field length, and (iv) accommodation in an x-ray hutch with its size typically $\sim 3$--4 m.
The last one is a practical aspect of x-ray experiments and limits the use of large bending magnets that were employed in previous experiments using lasers.
To satisfy these requirements, we have newly designed a small pulsed-magnet with a racetrack shape that has a field length of 20 cm.
In general, pulsed magnets are used for the study of material properties and suited to produce higher fields than that of superconducting magnets~\cite{pulmag, racetrack}.
Our magnet currently produces a peak field of up to 12 T at the magnet center, and we use multiple magnets to increase the total field length.
The details of the magnet structure, properties, and operation have been described in Ref.~\cite{our_mag}.
The magnet features high cooling-efficiency and small heating loss due to a low resistance coil, enabling highly repetitive operation with 0.2 Hz, whereas that of a conventional pulsed magnet is typically $\sim1$ mHz~\cite{pulmag}.
The time window associated with the pulse significantly reduces background counts at a detector, yielding a clean measurement for LSW searches.

\begin{figure}[b]
\includegraphics[angle=90,width=87mm]{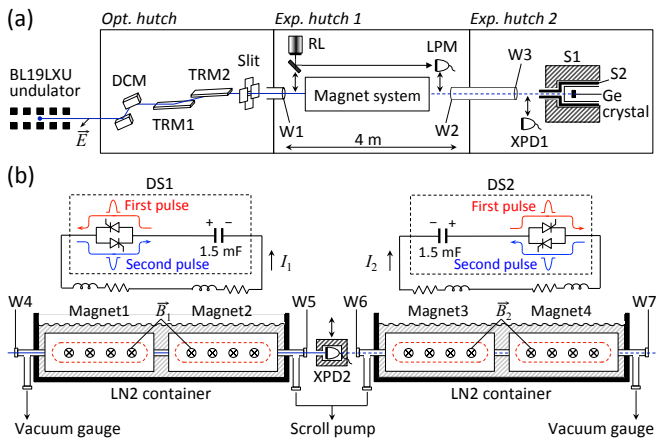}
\caption{
Schematic of the experimental setup.
(a) A layout of components in the optics hutch and in the two experimental hutches.
Retractable components are shown with a vertical arrow.
(b) A side view of the magnet system.
The magnets are placed so as to produce parallel fields with respect to the horizontal polarization of the x-ray beam.
Two identical discharge sections (DS1 and DS2) that are contained in the same hutch supply pulsed currents to the magnets by $LC$-discharge circuits.
\label{fig:setup}}
\end{figure}

Figure~\ref{fig:setup} shows a schematic of the setup at the BL19LXU beamline in SPring-8~\cite{bl19}.
The undulator has a length of 25 m and provides high intensity x rays.
The beams from the undulator are horizontally polarized and have the same time structure of electron bunches circulating the storage ring.
The bunch interval was 23.6 ns and is much smaller than the pulse duration of the magnetic field.
The undulator gap is tuned to produce x rays with an energy of $\omega=9.5$ keV.
The direct beam from the undualtor enters the optics hutch where fixed beamline-components have been installed to arrange the beam properties.
First, the beam is monochromatized with a double crystal monochromator (DCM) to a band width of $\Delta \omega/\omega=10^{-4}$, much smaller than the energy resolution of the detector.
Higher-order radiations are then removed by a pair of total reflection mirrors (TRMs).
Streaks from the mirrors are blocked at a four-jaw slit placed after the mirrors.
The beam size was measured to be 1.0 mm (H)$\times$0.5 mm (V) by scanning the slit.

\begin{figure}[b]
\includegraphics[clip,width=85mm]{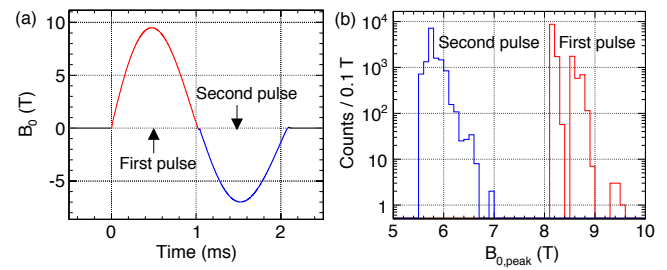}
\caption{
(a) Field shape of a shot with a charged voltage of 4.0 kV at the magnet center ($B_{0}$).
The first pulse has a peak field of 9.5 T while the second one reduces to 7.0 T due to the heat loss of coil resistance.
(b) Peak-field distribution of all 27,676 pulses during the run.
}
\label{fig:mag}
\end{figure}

The beam from the optics hutch enters the first experimental hutch where the magnet system is placed.
It consists of four racetrack magnets and their power supply with two identical discharge sections (DS1 and DS2).
The field length of a magnet is given by the straight section of the racetrack coil (shown with dashed lines in Fig.~\ref{fig:setup}(b)), that has a length of 200 mm along the beam axis.
The magnets produce parallel fields with respect to the polarization of x rays.
The beam pipe passing through the magnet has a diameter of 1/4 inch, much larger than the beam size.
To reduce the coil resistance, the magnets are placed in insulating containers and are cooled with liquid nitrogen.
As a circuit component, each magnet is regarded as a series of inductance ($\sim40$ $\mu$H) and resistance ($\sim25$ m$\Omega$).
Two magnets are connected in series and are supplied pulsed current from a charged capacitor of 1.5 mF (Nichicon, CCFI-652450HGW$\times$6) by applying an $LC$-discharge circuit.
Figure~\ref{fig:mag}(a) shows a typical field shape.
The shot consists of two successive pulses with each duration of about 1 ms.
The first pulse is triggered by a forward thyristor (DTI, T77P3000S12100$\times$4) while the second one by a reversed thyristor.
The same trigger pulse is divided by a pulse transformer (Nihon Pulse Industry, EX-B865) and is distributed to the thyristors to synchronize the upstream and downstream currents ($I_{1}$ and $I_{2}$).
The sum of the two currents ($I_{{\rm sum}}=I_{1}+I_{2}$) is read out with a current transformer (Pearson Electronics, Model 14239) and the waveform is recorded for each shot by a digitizer (NI PCI-6255).
After the second pulse, the energy lost by Joule heating in the magnets is additionally charged to the capacitors.

A small volume germanium detector (Canberra, GL0210) is placed in the next hutch to detect signal x rays.
The crystal has a diameter of 16 mm and a thickness of 10 mm.
The energy resolution and the detection efficiency, including the attenuation in its beryllium window, are measured with checking sources and obtained to be $\sigma = 93$ eV and $\epsilon=\left( 89\pm1 \right) \%$, respectively at 9.5 keV.
The end cap of the detector is shielded from environmental radiation by lead blocks with a thickness of 50 mm (S1).
Characteristic x rays from the lead and stray x rays of the beam are carefully removed with an inner shield of stainless steel (S2) attached to the end cap.
The shields reduce background events to a count rate of $\sim0.5$ mHz within a signal region of $\omega\pm2\sigma$.

Silicon PIN photodiodes (Hamamatsu, S3590-09) are used to measure x-ray flux.
One photodiode (XPD1) is inserted in front of the detector to measure the flux including the attenuation in upstream air and in vacuum windows (W1-W7).
Another photodiode (XPD2) is embedded in a lead block with an opening and placed at the central space between the two pairs of the magnets.
It serves as a beam dump while it also precisely monitors the flux to check for the photon loss due to misalignment.
XPD2 is placed on a slide rail and is retracted to off-axis position during the measurement with XPD1.
The pipe center of each magnet is first aligned using a 650-nm reference laser (RL), whose axis coincides with the x-ray beam, and its power meter (LPM) placed downstream of the magnets.
RL and LPM are then retracted to off-axis positions, and the x-ray flux is confirmed in front of the detector with XPD1.
The detector is also aligned with the same procedure.
After the alignment, the pipe ends of Magnet1 (Magnet3) and Magnet2 (Magnet4) are connected with a vacuum joint.
The beam path between W4 (W6) and W5 (W7) is evacuated with a scroll pump to a pressure less than 10 Pa.
Most of the other beam paths are also under vacuum to avoid x-ray attenuation in air.

The data-acquisition run was carried out in November 2015.
The magnets were operated with a pulse repetition of 0.2 Hz and charged voltages of 3.5--4.0 kV.
The total amount of heat generation in the four magnets was about 2 kW, requiring the supply of liquid nitrogen to the containers every 1.5 hours.
During the supply of liquid nitrogen, operation of the magnets was stopped, and the beam flux ($F$) at the detector was measured with XPD1.
The mean flux throughout the beam time was $\left( 3.0\pm0.1 \right) \times10^{13}$ ${\rm photons \cdot s^{-1}}$.
A total of 27,676 pulses was generated during two days of the net run time.
The peak-field distribution at the magnet center is shown in Fig.~\ref{fig:mag}(b).
The means are 8.3 T and 5.7 T for the first and second pulses, respectively.

\begin{figure}[t]
\includegraphics[width=80mm]{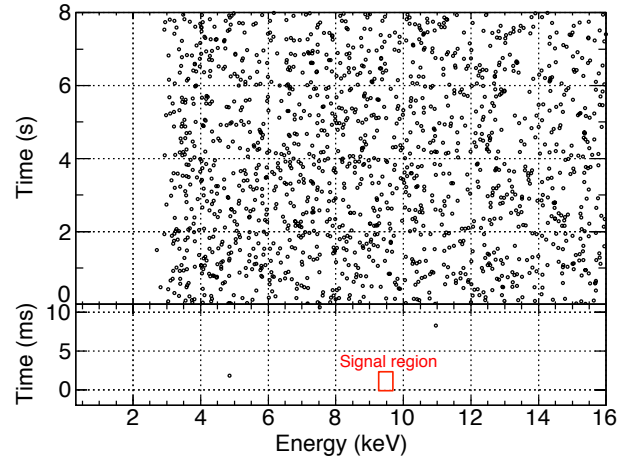}
\caption{
Top:
time-energy distribution of x rays measured with the germanium detector.
Each circle represents an event.
The horizontal axis is the time from the beginning of the first pulse.
The energy threshold was set around 3 keV.
Bottom: 
events around the field duration of 2.1 ms.
The signal region is shown around the beam energy $\pm 2 \sigma$ within the time window.
\label{fig:energy}}
\end{figure}

Signal candidates are characterized by the coincidence with pulsed fields and by an energy around $\omega$.
The timing of events is measured from the beginning of the first pulse.
Figure~\ref{fig:energy} shows the time-energy distribution of all events.
After a timing rejection of events outside the 2.1-ms window,
only one event is observed at 4.8 keV, hitting the detector at 1.8 ms.
The unsynchronized events are monotonously distributed as shown in Fig.~\ref{fig:energy}(top) and give an expected value for accidental counts within the time window as 0.96$\pm$0.62 counts below 16 keV.
This is consistent with the observed one event.
No coincident events are detected in the energy region of $\omega \pm 2 \sigma$.

The limit on $g_{a\gamma\gamma}$ is calculated from the null observation.
Since the field and its current change with time, the conversion probabilities for $\gamma \to a$ and $a\to \gamma$ also become time-dependent.
The expected signal count $N$ is expressed as
\begin{equation}
N = \epsilon \, \int_{0}^{t_{{\rm tot}}} \, P_{1}(t) \, P_{2}(t) \, F \, dt,
\label{eq:n}
\end{equation}
where $N=3.00$ at 95\% C.L. of a Poisson distribution, $P_{i}(t)$ the conversion probability of upstream $(i=1)$ and downstream $(i=2)$ regions, and $t_{{\rm tot}}$ the total pulse duration summed up over all shots.
The conversion probabilities are written as~\cite{lsw_calc}
\begin{equation}
P_{i}(t) = \frac{g_{a \gamma \gamma}^{2}}{4} \frac{\omega}{\sqrt{\omega^{2} - m_{a}^{2}}} \left| \int B_{i}(z, t)\,e^{iqz}\,dz \right| ^{2},
\label{eq:p}
\end{equation}
where $B_{i}(z, t)$ is the product of field map $B/I_{i}(z)$ and current $I_{{\rm sum}}(t)/2$, and $q=m_{a}^{2}/2\omega$ the momentum transfer to the field.
The field map of each magnet was measured along the beam direction $z$ at the pipe center by a calibrated pick-up coil.
It is expressed as a field-current ratio that is typically 0.60 ${\rm T\cdot kA^{-1}}$ at the magnet center.

\begin{table}[!t]
\caption{Summary of systematic uncertainties on $g_{a\gamma\gamma}$. }
\label{tab:systematic}
\begin{ruledtabular}
\small
\begin{tabular}{lc}
\multicolumn{1}{l}{Source} & Error (\%)       \\ \hline
Variation of x-ray flux $F$& $\pm0.1$ \\
Detection efficiency $\epsilon$& $\pm0.4$ \\
Accuracy of field-map measurement $B/I$& $\pm 0.9$ \\
Variation of field-current ratio $I_{{\rm sum}}$ vs $B$ & $\pm 1.2$ \\
Difference between $I_{1}$ and $I_{2}$ & $\pm0.3$ \\
Deviation of beam path from the pipe center& $+6.4$ \\
Total & $^{+6.5}_{-1.6}$ \\
\end{tabular}
\end{ruledtabular}
\end{table}

\begin{figure}[!b]
\includegraphics[clip,width=85mm]{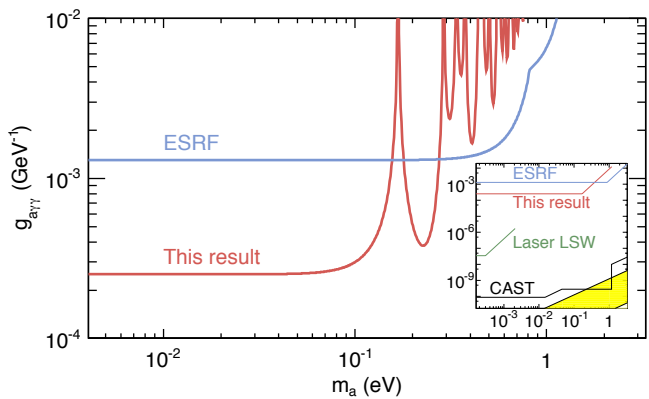}
\caption{
Upper limit on the ALP-two-photon coupling constant $g_{a\gamma\gamma}$ at 95\% C.L. as a function of the ALP mass.
The previous x-ray LSW limit obtained at the ESRF~\cite{esrf} is shown for comparison.
Inset: a large-scale view.
Previous laser LSW searches~\cite{alps, osqar} and solar axion searches (CAST~\cite{cast1, cast2, cast3, cast4}) are shown along with the band of invisible axion models~\cite{model1, model2, model3, model4}.
\label{fig:limit}}
\end{figure}

Systematic uncertainties are summarized in Table~\ref{tab:systematic}.
The accuracy of the field-map measurement is evaluated with another measurement by turning the pick-up coil by 180$^{\circ}$ from the original direction.
The variation of the field-current ratio is measured for the range of 5--9 T, covering most of the peak field values in the run.
Since the sum of the two currents $I_{{\rm sum}}(t)$ is measured during the run, each of them is measured one by one in additional measurements with the same setup to evaluate the contribution from their slight difference.
The largest uncertainty comes from the accuracy of the magnet alignment.
When the beam deviates from the pipe center, it may feel a weaker field than that at the center.
This contribution is evaluated with the worst path and calculated with a finite element simulation (ANSYS~\cite{ansys}) that includes all the 3D-geometrical information of the magnet.
It is conservatively estimated so that the beam paths of all magnets are assumed to be the worst one.
Since taking three events from the null observation is already quite conservative, an upper limit on $g_{a\gamma\gamma}$ at 95\% C.L. is calculated with $+1\sigma$ of the total systematic uncertainty and obtained to be
\begin{equation}
g_{a \gamma \gamma} < 2.51 \times 10^{-4}\;{\rm GeV^{-1}},
\label{eq:result_sys}
\end{equation}
below the ALP mass $\sim0.1$ eV, as shown in Fig.~\ref{fig:limit}.
Oscillation of the sensitivity appears for the heavier mass region because of the phase difference between photons and ALPs in Eq.~(\ref{eq:p}).
This effect becomes negligible for small ALP masses including the 20-meV region, and the result improves the previous x-ray limit by a factor of 5.2.
The relevant mass region scales to the square root of the photon energy.
Thus, it is higher by about two orders of magnitude compared to those probed by laser experiments (Fig.~\ref{fig:limit}, inset).

We are currently improving the magnet system to obtain a further gain on the sensitivity.
First, as demonstrated here, the total field length of our multi-magnet configuration can be flexibly changed.
This approach is thus suited to scale up the system by fabrication of many magnets~\cite{next_lsw}.
Another improvement can be envisaged by increasing the field strength that is currently restricted by the mechanical strength of a coil wound with a Cu wire.
Changing the wire material would provide a better mechanical strength.
For example, a peak field of 85.8 T has been obtained by a single solenoid wound with a Cu-Ag wire~\cite{kindo1, kindo2, kindo3}.
Besides, the system provides an efficient way to study nonlinear vacuum effects with large statistics~\cite{gamgam1, gamgam2, gamgam3, gamgam4, cm, bmv, pvlas1, pvlas2}.
While recent observation of neutron stars suggests the evidence for VMB~\cite{m7},
terrestrial measurements require a small gain on the current sensitivity~\cite{pvlas2}.
Since the signal birefringence increases in proportion to the square of the field strength,
high fields from pulsed magnets are advantageous~\cite{bmv, our_mag}.
In addition, the time variation of the field is essential to distinguish the signal from a large static birefringence of cavity mirrors~\cite{mirror1, mirror2}.
The novel system also provides various applications for studies on material properties under high-field conditions~\cite{rikken}.
Speaking only to SR facilities, many efforts have been made to technically combine a conventional pulsed magnet with the x-ray beam~\cite{narumi, frings, linden, matsuda, islam1, islam2, duc}.
Thus, transportability of the system would further increase the experimental opportunities.

In summary, we searched for pseudoscalar ALPs with a new x-ray LSW setup using highly repetitive racetrack-magnets that yield a clean measurement without background events.
An upper limit was imposed on the ALP-two-photon coupling constant that improves the one from a previous x-ray LSW search by a factor of 5.2.
The system provides a new method of probing vacuum under a high magnetic field with large statistics and various applications for studies on material properties.
The transportability of the system makes its field widely available to other experiments and facilities.

The authors thank S. Moriyama, M. Yokoyama, M. Nakahata, N. Saito, M. Ibe, Y. Narumi, Y. H. Matsuda, and T. Yamaji for their useful discussions. 
The experiment was performed at BL19LXU in SPring-8 with the approval of RIKEN (Proposal No. 20140024 and 20150010).
This work was also carried out under the Inter-university Cooperative Research Program of the Institute for Materials Research, Tohoku University (Proposal No. 14K0018 and 15K0080), and funded in part by JSPS KAKENHI (Grant No. 13J07172) and MEXT KAKENHI (Grant No. 26104701).


\begin{thebibliography}{00}
\bibitem{qed1}{
O. Halpern,
	Phys. Rev. {\bf 44}, 855 (1933).
}
\bibitem{gamgam1}{
F. Moulin \etal,
	Z. Phys. C {\bf 72}, 607 (1996).
}
\bibitem{gamgam2}{
D. Bernard \etal,
	Eur. Phys. J. D {\bf 10}, 141 (2000).
}
\bibitem{gamgam3}{
T. Inada \etal,
	Phys. Lett. B {\bf 732}, 356 (2014).
}
\bibitem{gamgam4}{
T. Yamaji \etal,
	Phys. Lett. B {\bf 763}, 454 (2016).
}
\bibitem{cm}{
C. Rizzo \etal,
	Int. Rev. Phys. Chem. {\bf 16}, 81 (1997).
}
\bibitem{pvlas1}{
F. Della Valle \etal,
	Phys. Rev. D. {\bf 90}, 092003 (2014).
}
\bibitem{pvlas2}{
F. Della Valle \etal,
	Eur. Phys. J. C {\bf 76}, 24 (2016).
}
\bibitem{bmv}{
A. Ca\`dene \etal,
	Eur. Phys. J. D {\bf 68}, 16 (2014).
}
\bibitem{qed2}{
B. De Tollis,
	Nuovo Cimento {\bf 32}, 757 (1964).
}
\bibitem{qed3}{
B. De Tollis,
	Nuovo Cimento {\bf 35}, 118 (1965).
}
\bibitem{primakov1}{
H. Primakov,
	Phys. Rev. {\bf 81}, 899 (1951).
}
\bibitem{primakov2}{
K. Van Bibber, N. R. Dagdeviren, S. E. Koonin, A. K. Kerman, and H. N. Nelson,
	Phys. Rev. Lett. {\bf 59}, 759 (1987).
}
\bibitem{atlas}{
ATLAS Collaboration,
	Phys. Lett. B {\bf 716}, 1 (2012).
}
\bibitem{cms}{
CMS Collaboration,
	Phys. Lett. B {\bf 716}, 30 (2012).
}
\bibitem{pdg}{
K. A. Olive \etal (Particle Data Group),
	Chin. Phys. C {\bf 38}, 090001 (2014).
}
\bibitem{axion1}{
R. D. Peccei and H. R. Quinn,
	Phys. Rev. Lett. {\bf 38}, 1440 (1977).
}
\bibitem{axion2}{
R. D. Peccei and Helen R. Quinn,
	Phys. Rev. D {\bf 16}, 1791 (1977).
}
\bibitem{cp1}{
C. A. Baker \etal,
	Phys. Rev. Lett. {\bf 97}, 131801 (2006).
}
\bibitem{cp2}{
S. Weinberg,
	Phys. Rev. Lett. {\bf 40}, 223 (1978).
}
\bibitem{cp3}{
F. Wilczek,
	Phys. Rev. Lett. {\bf 40}, 279 (1978).
}
\bibitem{susy}{
L. Covi, J. E. Kim, and L. Roszkowski,
	Phys. Rev. Lett {\bf 82}, 4180 (1999).
}
\bibitem{string1}{
P. Svr\v{c}ek and E. Witten,
	J. High Energy Phys. {\bf 06}, 051 (2006).
}
\bibitem{string2}{
M. Cicoli \etal,
	J. High Energy Phys. {\bf 12}, 146 (2012).
}
\bibitem{csm}{
K. A. Meissner and H. Nicolai,
	Phys. Lett. B {\bf 660}, 200 (2000).
}
\bibitem{ringwald}{
A. Ringwald,
	Dark Universe {\bf 1}, 116 (2012).
}
\bibitem{lsw_summary}{
J. Redondo and A. Ringwald,
	Comtemp. Phys. {\bf 52}, 211 (2011).
}
\bibitem{laser1}{
G. Ruoso \etal,
	Z. Phys. C {\bf 56}, 505 (1992).
}
\bibitem{laser2}{
R. Cameron \etal,
	Phys, Rev. D {\bf 47}, 3707 (1993).
}
\bibitem{pulse1}{
C. Robilliard \etal,
	Phys, Rev. Lett. {\bf 99}, 190403 (2007).
}
\bibitem{laser3}{
A. S. Chou \etal,
	Phys, Rev. Lett. {\bf 100}, 080402 (2008).
}
\bibitem{laser4}{
A. Afanasev \etal,
	Phys, Rev. Lett. {\bf 101}, 120401 (2008).
}
\bibitem{laser5}{
P. Pugnat \etal,
	Phys. Rev. D {\bf 78}, 092003 (2008).
}
\bibitem{pulse2}{
M. Fouche \etal,
	Phys, Rev. D {\bf 78}, 032013 (2008).
}
\bibitem{laser6}{
A. Afanasev \etal,
	Phys, Lett. B {\bf 679}, 317 (2009).
}
\bibitem{laser7}{
K. Ehret \etal,
	Mucl. Instrum. Meth. A {\bf 612}, 83 (2009).
}
\bibitem{alps}{
K. Ehret \etal,
	Phys. Lett. B {\bf 689}, 149 (2010).
}
\bibitem{osqar}{
R. Ballou \etal,
	Phys. Rev. D {\bf 92}, 092002 (2015).
}
\bibitem{model1}{
J. E. Kim,
	Phys. Lett. {\bf 43}, 103 (1979).
}
\bibitem{model2}{
M. A. Shifman \etal,
	Nucl. Phys. B {\bf 166}, 493 (1980).
}
\bibitem{model3}{
A. R. Zhitniskiy,
	Yad. Fiz. {\bf 31}, 497 (1980).
}
\bibitem{model4}{
M. Dine \etal,
	Phys. Lett. B {\bf 104}, 199 (1981).
}
\bibitem{lsw_summary2}{
P. W. Graham \etal,
	Ann. Rev. Nucl. Part. Sci. {\bf 65}, 485 (2015).
}
\bibitem{x_proposal1}{
R. Rabadan, A. Ringwald, and K. Sigurdson,
	Phys. Rev. Lett. {\bf 96}, 110407 (2006).
}
\bibitem{x_proposal2}{
A. G. Dias and G. Lugones,
	Phys. Lett. B {\bf 673}, 101 (2009).
}
\bibitem{esrf}{
R. Battesti \etal,
	Phys. Rev. Lett. {\bf 105}, 250405 (2010).
}
\bibitem{cast1}{
M. Arik \etal,
	Phys. Rev. Lett. {\bf 112}, 091302 (2014).
}
\bibitem{cast2}{
M. Arik \etal,
	Phys. Rev. Lett. {\bf 107}, 261302 (2011).
}
\bibitem{cast3}{
M. Arik \etal,
	JCAP {\bf 02}, 008 (2009).
}
\bibitem{cast4}{
S. Andriamonje \etal,
	JCAP {\bf 04}, 010 (2007).
}
\bibitem{sumico1}{
S. Moriyama \etal,
	Phys. Lett. B {\bf 434}, 147 (1998).
}
\bibitem{sumico2}{
Y. Inoue \etal,
	Phys. Lett. B {\bf 536}, 13 (2002).
}
\bibitem{sumico3}{
Y. Inoue \etal,
	Phys. Lett. B {\bf 668}, 93 (2008).
}
\bibitem{evade1}{
E. Mass\'{o} and J. Redondo,
	JCAP {\bf 09}, 015 (2005).
}
\bibitem{evade2}{
P. Jain and S. Mandal,
	Int. J. Mod. Phys. D {\bf 15}, 2005 (2006).
}
\bibitem{evade3}{
E. Mass\'{o} and J. Redondo,
	Phys. Rev. Lett. {\bf 97}, 151802 (2006).
}
\bibitem{evade4}{
P. Brax, C. van de Bruck, and A. C. Davis,
	Phys. Rev. Lett. {\bf 99}, 121103 (2007).
}
\bibitem{evade5}{
J. Jaeckel, E. Masso, J. Redondo, A. Ringwald, and F. Takahashi,
	Phys. Rev. D {\bf 75}, 013004 (2007).
}
\bibitem{evade6}{
A. K. Ganguly, P. Jain, S. Mandal, and S. Stokes,
	Phys. Rev. D {\bf 76}, 025026 (2007).
}
\bibitem{solar_mag}{
M. Ossendrijver,
	Astron. Astrophys. Rev. {\bf 11}, 287 (2003).
}
\bibitem{zioutas1}{
K. Zioutas \etal,
	New Jour. of Phys. {\bf 11}, 105020 (20049.
}
\bibitem{zioutas2}{
K. Zioutas \etal,
	arXiv:1003.2181 (proceedings of Patras workshop in 2009).
}
\bibitem{rusov1}{
V. D. Rusov \etal,
	arXiv:1401.3024.
}
\bibitem{rusov2}{
V. D. Rusov \etal,
	arXiv:1508.03836
}
\bibitem{pulmag}{
F. Herlach,
	Rep. Prog. Phys. {\bf 62}, 859 (1999).
}
\bibitem{racetrack}{
S. Batut \etal,
	IEEE Trans. Appl. Supercond. {\bf 18}, 600 (2008).
}
\bibitem{our_mag}{
T. Yamazaki \etal,
	Nucl. Instrum. Meth. A {\bf 833}, 122 (2016).
}
\bibitem{bl19}{
M. Yabashi \etal,
	Nucl. Instrum. Meth. A {\bf 467–-468}, 678 (2001).
}
\bibitem{lsw_calc}{
G. Raffelt and L. Stodolsky,
	Phys. Rev. D {\bf 37}, 1237 (1988).
}
\bibitem{ansys}{
ANSYS Multiphysics release 15.0, ANSYS Inc.,
	http://www.ansys.com
}
\bibitem{next_lsw}{
R. Bahre \etal,
	JINST {\bf 8}, T09001 (2013).
}
\bibitem{kindo1}{
K. Kindo,
	J. Phys. Conf. Ser. {\bf 51}, 522 (2006).
}
\bibitem{kindo2}{
K. Kindo \etal,
	J. Low Temp. Phys. {\bf 159}, 381 (2010).
}
\bibitem{kindo3}{
A. Matsuo \etal,
	proceedings of 10th International Conference on Research in High Magnetic Fields p134 (2012).
}
\bibitem{m7}{
R. P. Magnani \etal,
	arXiv:1610.08323.
}
\bibitem{mirror1}{
G. Bialolenker \etal,
	Appl. Phys. B {\bf 68}, 703 (1999).
}
\bibitem{mirror2}{
F. Bielsa \etal,
	Appl. Phys. B {\bf 97}, 457 (2009).
}
\bibitem{rikken}{
B. A. van Tiggelen and G. L. J. A. Rikken,
	Optical Properties of Nanostructured Random Media p. 275.
}
\bibitem{narumi}{
Y. Narumi \etal,
	J. Synchrotron Rad. {\bf 13}, 271 (2006).
}
\bibitem{frings}{
P. Frings \etal,
	Rev. Sci. Instrum. {\bf 77}, 063903 (2006).
}
\bibitem{linden}{
P. J. E. M. van der Linden \etal,
	Rev. Sci. Instrum. {\bf 79}, 075104 (2008).
}
\bibitem{matsuda}{
Y. H. Matsuda \etal,
	Phy. Rev. Lett. {\bf 103}, 046402 (2009).
}
\bibitem{islam1}{
Z. Islam \etal,
	Rev. Sci. Instrum. {\bf 80}, 113902 (2009).
}
\bibitem{islam2}{
Z. Islam \etal,
	Rev. Sci. Instrum. {\bf 83}, 035101 (2012).
}
\bibitem{duc}{
F. Duc \etal,
	Rev. Sci. Instrum. {\bf 85}, 053905 (2014).
}

\end{thebibliography}
\end{document}